# Design and Simulation of a high order mode cavity bunch length monitor

Jiang Guo (郭江)    ZeRan Zhou (周泽然)    Qing Luo (罗箐)

National Synchrotron Radiation Laboratory, University of Science and Technology of China, Hefei 230029, China

**Abstract:** A new bunch length measurement method based on high order mode cavity was proposed. Operating the harmonic cavity at mode TM0n0 so that its radius could be chosen, in order to break the limitation of beam pipe radius. A two-cavity bunch length monitor for linac of positron source was designed. Operating frequency selection for different bunch time structure was discussed and calculation formula of bunch length was deducted. Fundamental harmonic cavity resonates at 2.856 GHz with mode TM010. Fifth harmonic cavity resonates at 14.28 GHz (fifth harmonic of the linac fundamental frequency 2.856 GHz) with mode TM020, which could provide larger radius. Each cavity equipped with a filter to suppress unwanted signal. A simulation measurement was conducted in CST Particle Studio for beam current from 100-300mA, bunch length from 5-10ps, calculation results shows a fairly high accuracy (better than 3%). Several cases were discussed.

**1.Introduction**

  High performance linacs are used widely in FEL facilities, while linac-based positron source shows exciting potential in positron annihilation techniques. Both kinds of linacs could use photocathode rf guns as electron source to get short bunch. A high brightness photon injector (HLS High Brightness Injector) was built for Hefei Light Source to develop advanced accelerator technologies, based on the past experience in slow positron beam, NSRL has plan to build a fast positron beamline for deep tiny flaw detection [1]. Measuring and controlling the electron bunch length are crucial to linacs , therefore effective bunch length measurement methods are required. In this paper a new monitor designed for linacs using resonate cavities is discussed. This new method can be very useful for future electron and positron sources.

  Linac-based positron source has a bunch length ranges from 5 to 10 ps. Electronics methods such as stripline connecting to a digital sampling oscilloscope can be used to non-invasively obtain the bunch length, but these devices are not fast enough, with minimum rise time detection about 200ps[2]. A RF deflecting cavity can be used to measure shorter bunch length (few ps), but it's an invasive method[3]. It's very desirable to have a compact, real time and non-invasive bunch length monitor, particularly one that can detect bunch length about 10ps. Such a device can be used to accomplish a number of important tasks, including monitoring RF phase stability, setting the RF amplitude of bunching cavity, and detecting bunch length growth due to space charge forces[4].

  Cavity bunch length monitor could non-invasively provide signal with large amplitude and high signal-to-noise ratio,its measurement accuracy could increase with the development of electronics. This method has attracted a lot of attention recently, shows more development potential[5,6,7]. In China the method was first used by High Energy Institute to observe the bunch length in their linac injectors[8]. The radius of fifth harmonic cavity is so close to the beam pipe that it can be barely regards as a cavity. This could increase error to the final result and prevent the possibility to utilize higher order harmonic (sixth or higher) cavity, which have a smaller radius, but could improve measurement accuracy. Linac of proposed positron source for NSRL has a beam pipe radius 5mm, harmonic cavity radius in conventional method is almost equal to beam pipe. In this paper, a new bunch length measurement method based on high order mode cavity was proposed. Operating the harmonic cavity at mode $TM_{0n0}$ so that it's radius could be chosen, in order to break the limitation of beam pipe radius. A two-cavity bunch length monitor for linac of positron source was designed. Fifth harmonic cavity was resonated at frequency of mode $TM_{020}$, which could have a radius bigger than the cavity operating at $TM_{010}$. Each cavity equipped with a filter to suppress unwanted signal. Virtual beam with realistic parameters was used in computer simulation and several cases were discussed.

**2. Theoretical deduction and frequency selection**

**2.1 Theoretical deduction**

An amplitude ratio of two specified frequency components in the beam spectrum contains the information of bunch length. Several axially symmetric mode $TM_{0N0}$ can be excited after the beam pass through the cavity. Using cavities as pick-up, their response to the wakefield of beam passage gives the components of the beam spectrum at the two specified frequencies, from which the bunch length can be calculated.

A periodically bunched Gaussian electron beam was considered, each bunch include N electrons, spaced by time interval T (repetition frequency $\omega_0$), with an identical bunch length $\sigma_\tau$. The time-varying bunch current can be expressed by

$$I_{b-G}(t) = \frac{eN}{\sqrt{2\pi}\sigma_\tau} \exp(-\frac{t^2}{2\sigma_\tau^2}) \tag{1}$$

Which can be expressed by a Fourier series

$$I_b(t) = \frac{eN}{T} + \sum_{m=1}^{\infty} I_m \cos(m\omega_0 t) \tag{2}$$

The constant term $I_0 = \frac{eN}{T}$ represents the DC offset, successive terms represent contributions at integer multiples of $\omega_0$, $I_m$ is the amplitude of m-th harmonic current.

$$I_m = I_0 \exp(-\frac{m^2 \omega_0^2 \sigma_\tau^2}{2}) \tag{3}$$

For a cavity resonating at m-th harmonic of beam repetition frequency $\omega_0$, the beam induced power should be

$$P_m = I_m^2 R_m = \left[I_0 \exp(-\frac{m^2 \omega_0^2 \sigma_\tau^2}{2})\right]^2 R_m \tag{4}$$

Rm represents shunt impedance of cavity. For m=1, $\sigma_\tau$ about 10ps, $\exp(-\frac{m^2 \omega_0^2 \sigma_\tau^2}{2}) \approx 1$, therefore $P_1 \approx I_0^2 R_1$, obviously, the power of fundamental harmonic cavity is mainly decided by beam current and slightly effected by bunch length. For higher harmonic number (like m=5), bunch length and current both become important to output power. Bunch length could be calculated if beam current obtained from fundamental harmonic cavity and substituted to the power expression of fifth harmonic cavity. Substitute m=5 and $\omega_0 = 1.794 * 10^{10} \ rad/s$, the bunch length could be expressed by equation (5)

$$\sigma_\tau = 15.76 * \sqrt{\ln\sqrt{\frac{P_1 R_5}{P_5 R_1}}}\ (ps) \tag{5}$$

In conventional method operating mode of cavity is $TM_{010}$, so the radius of the harmonic cavity have to be very small in order to make the frequency of mode $TM_{010}$ to reach several times of repetition frequency. In the case of short bunch, cavity radius could be smaller than the beam pipe radius because of the high operating frequency, the method couldn't be used. In this paper, harmonic cavity operating at mode $TM_{0n0}$ was proposed, for a given frequency, cavity radius could be changed by value n of it's operating mode $TM_{0n0}$. The radius R could be expressed as

$$R = \frac{c}{2\pi}\frac{u_{0n}}{f_0} \tag{5}$$

Where $u_{0n}$ represent $n^{th}$ root of 0 order Bessel functions. Therefore adjust operating mode in the high frequency situation could ensure cavity radius bigger than beam pipe. A two-cavity bunch length monitor for linac-based positron source was designed. Fundamental harmonic cavity resonates at 2.856 GHz with mode $TM_{010}$. Fifth harmonic cavity resonates at 14.28 GHz with mode $TM_{020}$, the cavity radius is 19mm, bigger than beam pipe 5mm. when operating at $TM_{010}$, cavity radius is 6mm, almost equal with beam pipe.

**2.2 Operating frequency of two cavity**

The $\sigma_\tau$ measurement should be independent from the charge distribution in the bunch. Gaussian electron bunch was considered in situation above and in the real case time structure of the bunch is unknown, usually between ideal Gaussian and ideal rectangular. In order to measure the bunch length of different time structure precisely, cavity should avoid operating at high order harmonic of $\omega_0$, as a result of a huge difference in the high frequency domain in the spectrum of Gaussian beam and rectangular beam[9].

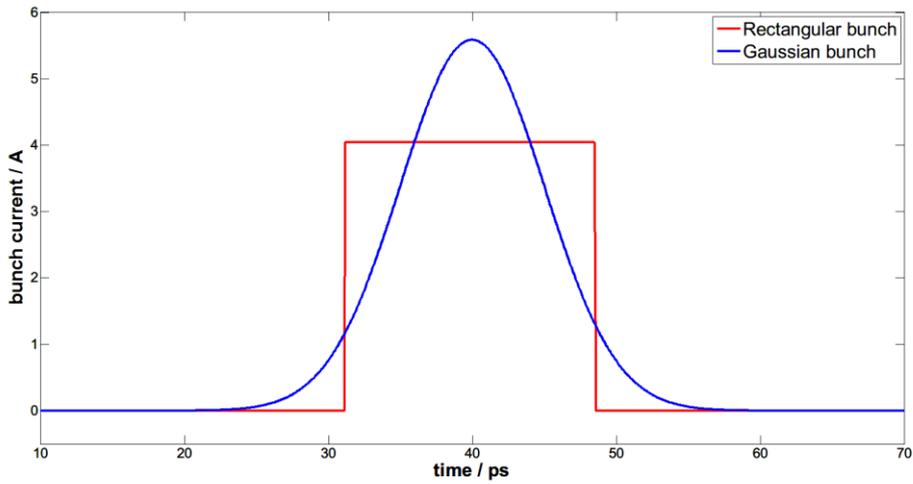

Figure 1. Time structure of Gaussian and rectangular bunch with identical bunch length $\sigma_\tau = 5ps$ and bunch current q=70.03pC.

As a comparison of equation (1), the time-varying bunch current of rectangular bunch with length $\sigma_\tau$ can be expressed as

$$I_{b-R}(t) = \begin{cases} \frac{1}{2\sqrt{3}\sigma_\tau} & |t| \leq \sqrt{3}\sigma_\tau \\ 0 & |t| > \sqrt{3}\sigma_\tau \end{cases} \quad (5)$$

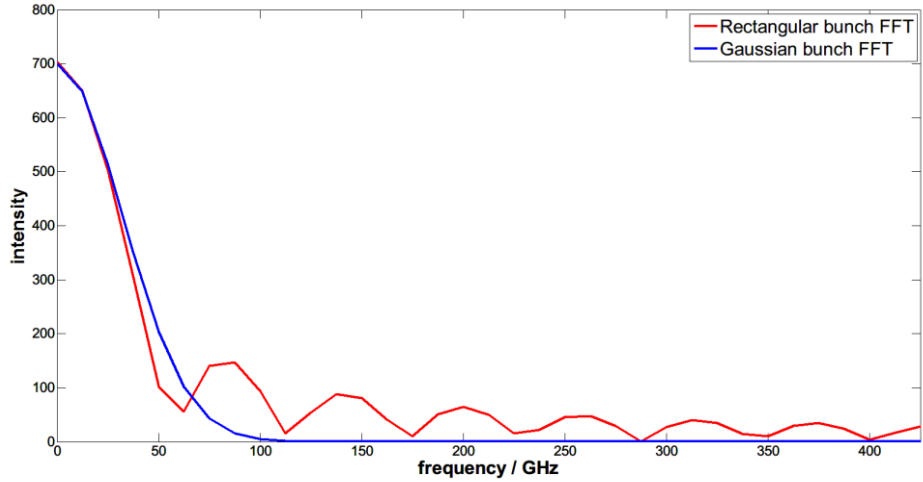

Figure 2. Fast Fourier Transformation of Gaussian and rectangular bunch

Figure 1 compares the time structure of Gaussian and rectangular bunch with identical length $\sigma_\tau = 5\text{ps}$. Figure 2 compares their spectrum. It can be observed that two spectra coincide in low frequency area and separate in high frequency area. Further study suggests that difference emerges after the point of 16 GHz which spectral intensity is 60% of its peak value at f=0. Therefore the operating frequency should be lower than 16GHz. For linac fundamental frequency $f_0 = 2.856\text{ GHz}$, fifth harmonic of $\omega_0$ is 14.28 GHz, just meet the criteria. So the fundamental harmonic cavity resonates at 2.856 GHz with mode $TM_{010}$, while the high order harmonic cavity resonates at 14.28 GHz with mode $TM_{020}$.

## 3. Design of cavity and filter

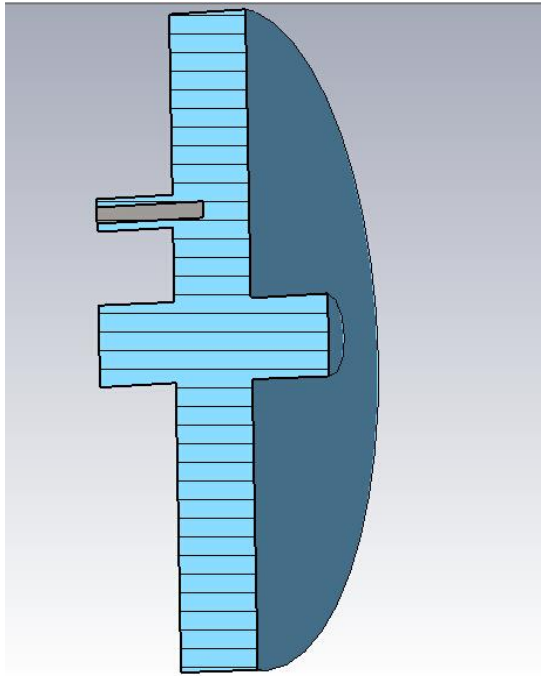

Figure 3. Model for simulation of fundamental harmonic cavity

**3.1 Design of fundamental harmonic cavity**

Ordinary pill-box cavity was used while its radius was adjusted to exactly resonate at 2.856 GHz with operating mode $TM_{010}$. Coaxial antenna was penetrated to couple out the signals. The frequency of $TM_{040}$ is 14.00GHz, close to fifth harmonic frequency 14.28 GHz, the frequency of $TM_{080}$ is 28.91 GHz. close to tenth harmonic frequency 28.56 GHz. As a result, both of these two higher order modes could be excited, though would be very weak. Adjust the position andpenetrate depth to avoid coupling of high order modes and obtain suitable $Q_e$.

**3.2 Design of fifth harmonic cavity**

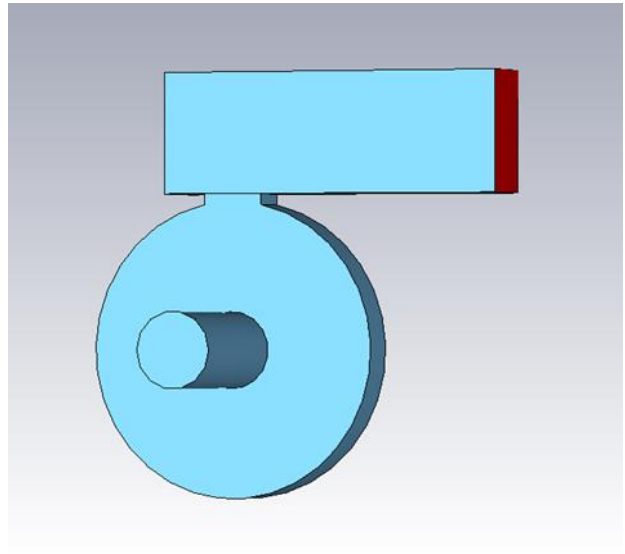

Figure 4. Model for simulation of fifth harmonic cavity

The radius of ordinary pill-box cavity was adjusted to exactly resonate at 14.28GHz with operating mode $TM_{020}$. Optimum radius is 19.27 mm, compared with cavity operating at $TM_{010}$, which has an optimum radius of 6 mm, slightly larger than the beam pipe, therefore the high order mode cavity could help to reduce the measurement error and make the fabrication easier. Coupling slot is used to couple out signals to waveguide, adjust waveguide size to transmit 14.28 GHz signal and suppress possible low frequency signal. Standard waveguide BJ-140 (15.799*7.899 mm) was used for convenience, which have a cutoff frequency of 11.9GHz.

**3.3 Design of Coaxial line filter for fundamental harmonic cavity**

Preliminary simulation result shows several high frequency component mixed in output signal of both cavity. Measurement of bunch length is decided by amplitude of output signal, which is affected by these high frequency component. So it's necessary to add a filter to suppress these unwanted signal. This could help to optimize output signal, and hard filtering based on microwave devices could greatly release the burden of following signal processing circuit.

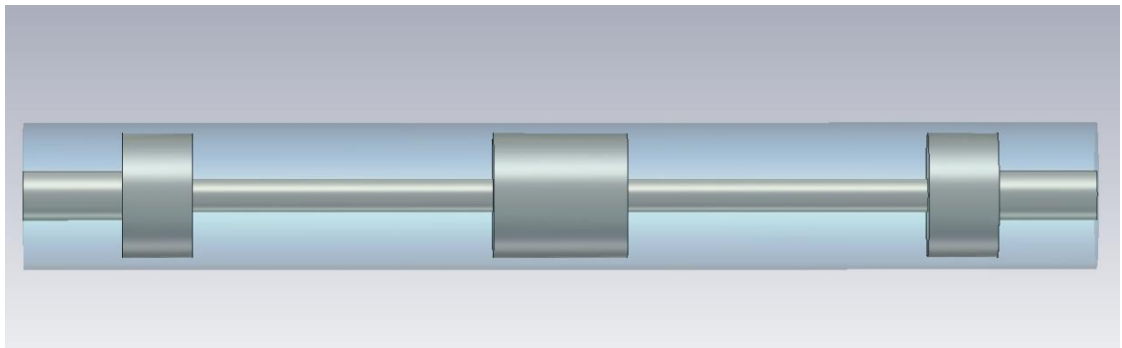

Figure 5. Model for simulation of low pass coaxial filter. Inner conductor of coaxial line is made by different radius to formulate high Impedance section and low impedance section

A 5th order low pass filter based on coaxial line is designed because coaxial probe is used to couple out the signal. Low pass Chebyshev prototype was considered to determine its equivalent circuit, which is composed of a number of inductors and capacitors. The inner conductor of coaxial line was made by different radius to formulate high Impedance section and low impedance section. The shift between these sections could provide adequate inductors and capacitors. Figure 5 shows the simulation model of coaxial line filter.

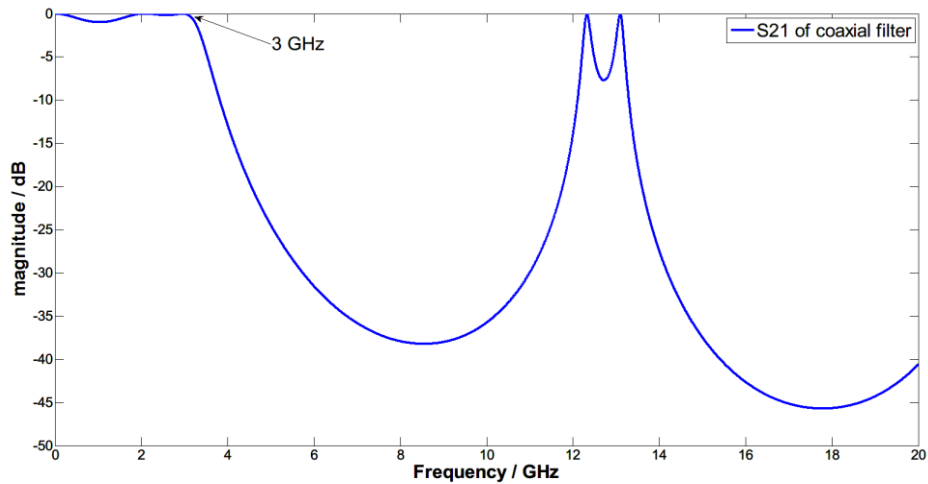

Figure 6. S21 of coaxial line low pass fiter

The S21 of filter shows in figure 6. Cutoff frequency is 3GHz considering 2.856 GHz as main frequency. Insertion loss of passband is well below -1dB. The parasitic passband appear in 12~14GHz, it's adjusted to avoid frequency of unwanted signal. Attenuation to main unwanted high frequency (14.28 GHz) signal is greater than 30dB.

**3.4 Design of waveguide filter for fifth harmonic cavity**

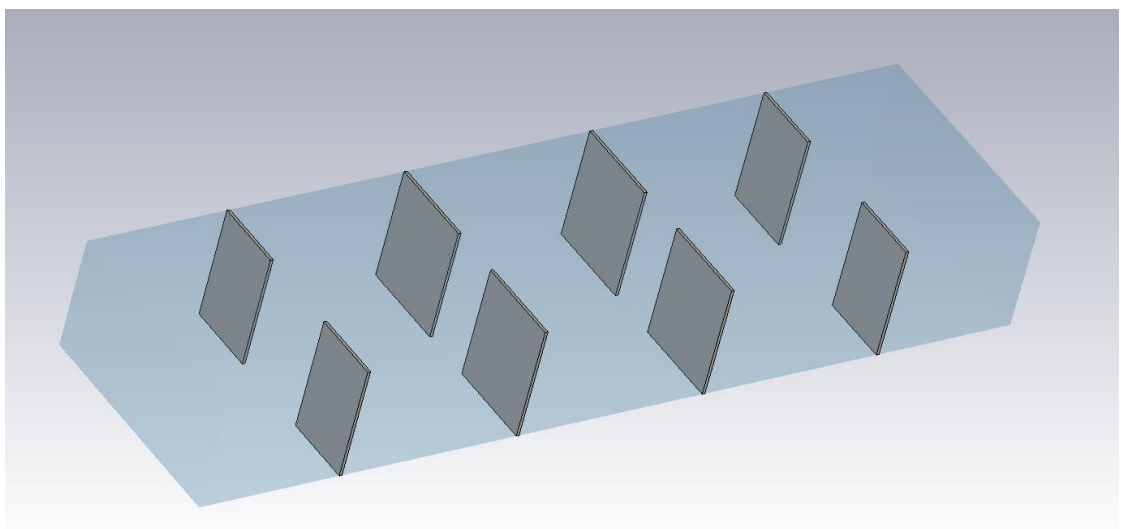

Figure7. Model for simulation of waveguide bandpass filter. Eight grey squares are metal iris

A 4th order waveguide iris bandpass filter is designed. Frequency shift based on low pass Butterworth prototype was deducted to determine its equivalent circuit[10]. Metal irises were loaded in the standard waveguide BJ-140. Simulation model of iris-loaded waveguide filter shows in figure 7. Inductance iris and its adjacent two waveguide segments worked as impedance inverter, while half wavelength waveguide segment acts as parallel resonant circuit.

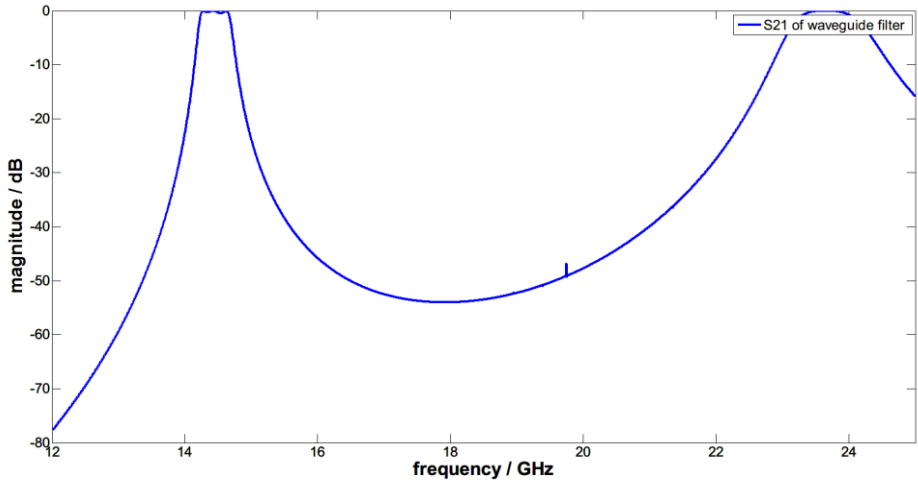

Figure 8. S21 of waveguide bandpass fiter

The S21 of filter shows in figure 8. The size and spacing of iris were adjusted to meet the goal of bandpass filter: center frequency of passband 14.28GHz, passband bandwidth 500MHz, Insertion loss of passband is well below -1dB, Attenuation to main high frequency 19.99GHz is greater than 30dB.

**4. Simulation and bunch length measurement**

The cavity bunch length monitor above was loaded with virtual beam to give a measurement simulation in CST Particle Studio. Cable attenuation coefficient was not considered. Repetition rate of virtual beam is 2.856 GHz.

**4.1 Fundamental harmonic cavity as a current monitor**

For fundamental harmonic cavity, $P_1 \approx I_0^2 R_1$, the change of bunch length could slightly affects the amplitude of output signal, which could be proved by simulation: change the bunch length from 5ps to 10ps while beam current is 300mA, the amplitude of output signal changed by 4%, output signal amplitude of fundamental harmonic cavity is shown in table 1. therfore it could work as a current monitor: Substitute shunt impedance $R_1$ and power $P_1$ of output signal, current $I_0$ could be calculated.

Table 1. Table 2. fundamental harmonic cavity measurement results of different current

| Bunch length / ps | Output signal amplitude / dB |
|---|---|
| 5 | 74.20 |
| 6 | 72.95 |
| 7 | 72.14 |
| 8 | 71.51 |
| 9 | 71.14 |
| 10 | 70.76 |

Submitted to Chinese Physics C

Change the current from 100mA to 300mA while bunch length is 10ps, output signal of fundamental harmonic cavity is shown in figure 9. Vertical coordinate represents normalized electromagnetic field signal intensity. The signal waveform is composed by two section: ascending section and equilibrium section. Cavity, as a high impedance structure, would generate power $P_1$ while beam passed. The energy of electromagnetic field gradually increase. The power coupling out $P_{out}$ is proportional to field energy by coefficient $1/Q_e$. In the ascending section, $P_1 > P_{out}$, field energy and $P_{out}$ grow synchronously. Until equilibrium is reached: field energy stabilize and $P_1 = P_{out}$. The calculation results is shown in table 2 compared with current.

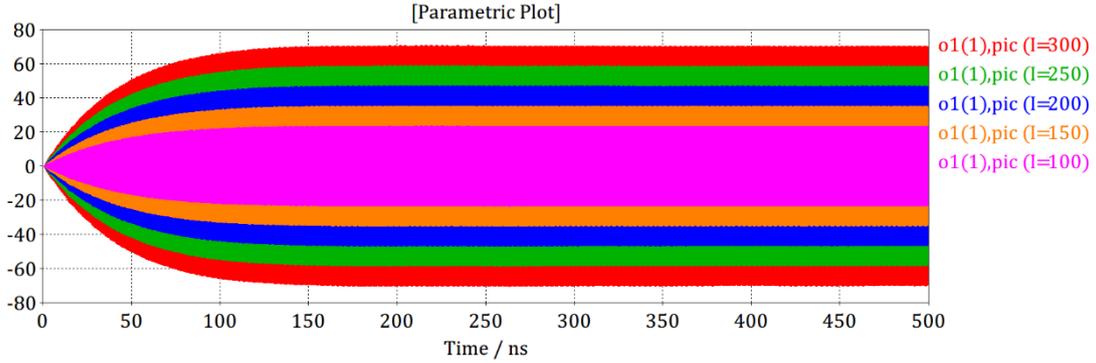

Figure 9. output signal of fundamental harmonic cavity（current from 100mA to 300mA）

| Beam current / mA | Measurement results / mA | Error |
| --- | --- | --- |
| 100 | 100.3 | 0.30% |
| 150 | 150.5 | 0.33% |
| 200 | 199.7 | 0.15% |
| 250 | 250.1 | 0.04% |
| 300 | 299.8 | 0.07% |

Table 2. fundamental harmonic cavity measurement results of different current

The spectrum of output signal after filter is a individual peak which center frequency is 2.856 GHz, therefore, amplitude of output signal could accurately reflect the change of beam current.

**4.2 Fifth harmonic cavity as a bunch length monitor**

Change the bunch length from 5 ps to 10 ps while current is 200mA, output signal of fifth harmonic cavity is shown in figure 10. Vertical coordinate represents normalized electromagnetic field signal intensity. Spectrum of output signal after filter is a individual peak which center frequency is 14.28GHz. substitute power of output signal $P_1$ and $P_5$ to equation (5), bunch length could be solved. The calculation results is shown in table 3 compared with theoretical bunch

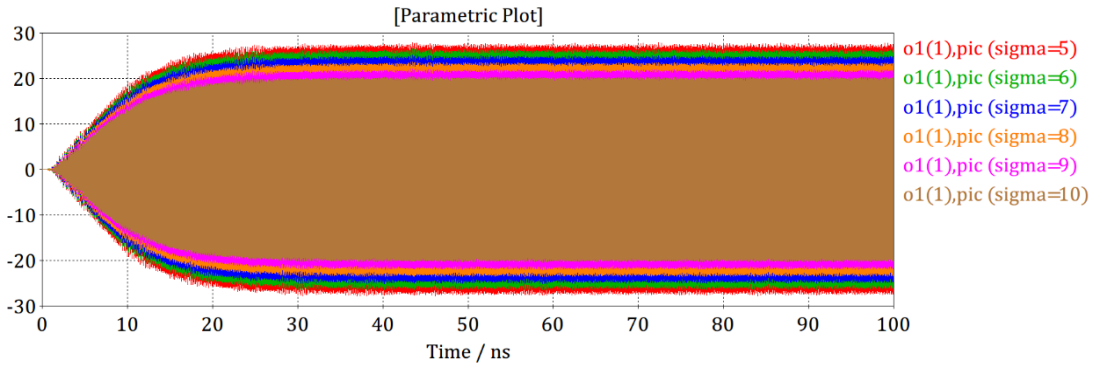

Figure 10. output signal of fifth harmonic cavity (bunch length from 5ps to 10ps)

length.

| Bunch length / ps | Measurement results / ps | Error |
|---|---|---|
| 5 | 4.87 | 2.60% |
| 6 | 5.95 | 0.80% |
| 7 | 7.01 | 0.14% |
| 8 | 8.04 | 0.50% |
| 9 | 9.06 | 0.67% |
| 10 | 10.07 | 0.70% |

Table 3. fifth harmonic cavity measurement results of different bunch length

From the calculation results, for all possible bunch length of linac of positron source (5~10ps), simulation measurement provides a fairly high accuracy (better than 3%). Electronic noise and nonlinearity were not considered in the simulation. Besides, power measurement could achieve very high accuracy in the simulation software, which is not possible in the real case.

NSRL has done signal processing for beam instrument in the past[11]. There are two ways to deal with the signal from the bunch length monitor. First, a microwave power meter could be used to directly measure $P_1$ and $P_5$. Power meter AV2436 (frequency range 10MHz-40GHz) could meet the requirements. Second, signal processing circuit with a down-converter module could be considered, mixing the signal down to an intermediate frequency (IF) within one or more stages. With an ADC to convert it to digital signal for further calculation.

**5. conclusion**

A new bunch length measurement method based on high order mode cavity was proposed. Operating the harmonic cavity at mode $TM_{0n0}$ so that its radius could be chosen, in order to break the limitation of beam pipe radius. A two-cavity bunch length monitor for linac-based positron source was designed. Fifth harmonic cavity with high order mode ($TM_{020}$) could provide larger radius for easier fabrication, and error from cavity too close to the beam pipe could also be avoided. Simulation measurement include all possible bunch length in the linac of positron source, calculation results of simulation shows a fairly high accuracy (better than 3%).